\documentstyle[11pt,paspconf,epsf,natbib209]{article}

\def\refname{References}

\def\rb{r_{\rm b}}
\def\mub{\mu_{\rm b}}
\def\rd{r_{\rm d}}
\def\mud{\mu_{\rm d}}
\def\magarc{\rm mag\ arcsec^{-2}}
\def\etal{{et al.}\ }
\def\kms{\rm km\,s^{-1}}
\def\ion#1#2 {#1\,{\small\rm{#2}}\ }
\def\arcdeg{\hbox{$^\circ$}}
\def\plotahb#1#2{\centering \leavevmode
    \epsfxsize=#1 \epsfbox{#2}}

\begin{document}

\title{Modeling the Mass Distribution in Spiral Galaxies}
\author{Adrick H. Broeils}
\affil{Stockholm Observatory, S-133 36 Saltsj\"obaden, Sweden}
\author{St\'ephane Courteau}
\affil{KPNO/NOAO, P.O. Box 26732, Tucson, Arizona 85726, USA}

\begin{abstract}
We use deep $r$-band photometry and H$\alpha$ rotation curves for a sample of
290 late-type spirals to model their mass distribution within the optical
radius.  We examine luminosity profile decompositions into bulge and disk
carefully and confirm that bulge light is best modeled by a seeing-convolved
exponential profile.  The optical rotation curves are well-reproduced with a
combination of bulge and ``maximum'' disk components only.  No dark halo is
needed.  The disk mass-to-light ratios ($M/L$'s) correlate with the ``size''
of galaxies, as measured by mass, luminosity, or disk scale length.
Correcting for this scale effect yields a narrow distribution of intrinsic
$M/L$'s for this galaxy population. By combining these models with \ion{H}{I}
data for other samples, we confirm that the luminous mass fraction increases
with galaxy ``size''.
\end{abstract}

\section{Introduction}

In order to improve our knowledge of the distribution and amount of luminous
and dark matter (DM) in spiral galaxies, one has to construct mass models to
reproduce the observed rotation curves (RCs) in detail (e.g. see review by van
Albada in this volume). This can only be done by using extended \ion{H}{I}
RCs which sample the matter distribution far beyond the stellar disk. A
troubling feature of these models is, however, the unknown scaling of the
light distribution to a mass distribution in the form of a free $M/L$ ratio.
This uncertainty limits us to determining only the relative contributions of
disk, bulge and dark components to the gravitational potential by making
certain assumptions for the $M/L$ ratios of the luminous components.  The most
widely used alternative is to maximize the contributions of these components
to the observed RC (``maximum disk hypothesis''). One of the arguments for
using a maximum disk is that optical RCs are very well reproduced by a
combination of luminous components only. New data for the Milky Way also seem
to be in agreement with a maximum disk solution with an $M/L$ typical for
Sb--Sc galaxies \cite[]{Sackett96}. In this paper we investigate the
distribution of disk $M/L$ ratios obtained from optical RCs (without the use
of a DM component), and try to determine the width of the intrinsic
(i.e. ``scale''-free) $M/L$ distribution.

\begin{figure}
\plotone{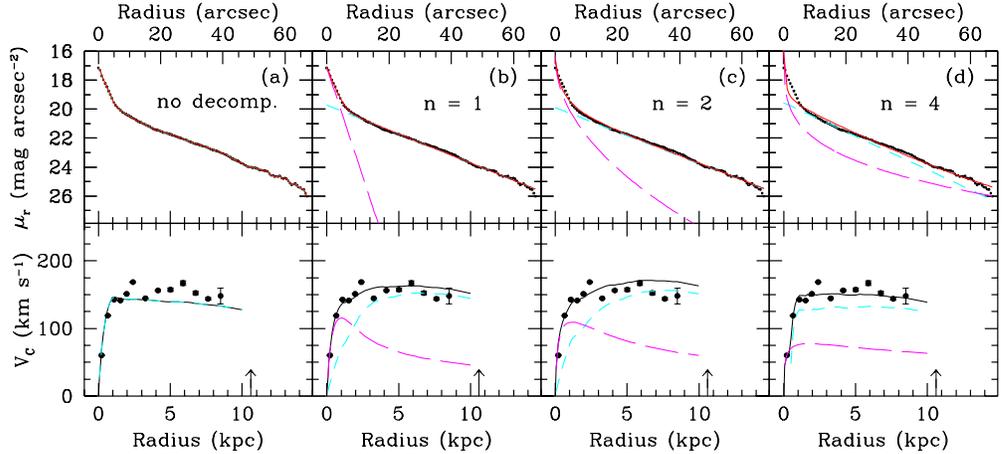}
\caption{Examples of bulge-disk decompositions (top panels) using S\'ersic
bulge laws with $n = 1$, 2, and 4 (long-dash) $+$ exponential disk (short
dash) and the corresponding mass decompositions (bottom panels) for UGC~9097
($b-d$).  The SB fit is generally poorer with greater $n$, even for galaxies
with substantial bulges like this one. A model without B/D decomposition fails
to reproduce to observed RC ($a$).  The arrow shows the location of R$_{24}$.
\label{fig_sub_sb_rc} }
\end{figure}

\section{Sample}
We have used data from the large collection of deep $r$-band photometry and
H$\alpha$ RCs of Sb and Sc galaxies by
\cite{Courteau92,Courteau96b,Courteau97}.  The sample was originally selected
for Tully-Fisher mapping of peculiar velocities in the Northern Hemisphere.
It includes Sb--Sc galaxies from the UGC catalog with inclinations between
$50\arcdeg < i < 80\arcdeg$, which makes this sample well suited for mass
modeling studies.  From the original published collection of 350 spirals, we
selected 290 galaxies with data of suitable quality to fit mass models to the
optical RCs.  The available data base is briefly described below; more
detailed information about the photometric and spectroscopic observations can
be found in \cite[]{Courteau96b} and \cite[]{Courteau97}.

The one-dimensional $r$-band surface brightness (SB) profiles were obtained
with 1m-class telescopes at Lick, Palomar, and Kitt Peak observatories.
Repeat observations taken for almost half of the total observed sample and
comparisons with the $r$-band photometry of \cite{Kent84,Kent86} and
\cite{Willick91} both indicate that the photometry is reliable down to 26
$r$-$\magarc$ and that the total $r$-band magnitudes are accurate to $\pm
0\fm08$.

Spectroscopic observations were made with the UV Schmidt spectrograph at the
Lick 3m telescope.  Exposure times were typically 1800\,s which yields a
sampling of the H$\alpha$ RCs out to about one optical radius ($R_{24}$,
defined at the 24 $r$-$\magarc$ isophotal level).  For the purpose of mass
modeling, the observed RCs were folded and averaged about their centers.
After inspection of each rotation curve, a transition radius was determined
just beyond the central solid-body rise.  The averaged, folded RCs were then
re-sampled at intervals of 2\arcsec\ (average seeing of all observations)
within the transition radius; beyond that, the bins were doubled in size to
improve the S/N in the fainter regions of the disk.  Line-of-sight velocities
were deprojected using disk inclinations determined from the photometry.  A
correction for redshift broadening was also applied.  Seventy-six galaxies
were observed more than once and their folded RCs were combined by taking the
weighted average velocity within each bin.

We also retrieved IRAS $60\,\micron$ fluxes $f_{60}$ for 189 sample galaxies
from the IRAS Faint Source Catalog through NED (NASA/IPAC Extragalactic
Database).  These provide an estimate of the star formation rate and will be
used in our discussion in \S~\ref{sec_ml} The conversion from observed
(\arcsec) to physical (kpc) units uses heliocentric redshifts and H$_0 = 70
\,\kms\,{\rm Mpc}$.

\section{Luminosity Decompositions}

\subsection{B/D decomposition technique}

It has become clear that one must derive bulge and disk parameters
simultaneously in order to obtain reliable decompositions \cite[]{Kormendy77a}. 
The method we adopt is to fit model functions to the bulge and disk profiles (in
the magnitude regime) using a non-linear $\chi^2$ minimalization routine.  The
sum of disk and bulge model functions were convolved with a Gaussian-shaped
PSF (with the appropriate seeing FWHM) before it was fitted to the observed
profile.

A fundamental aspect of the decompositions is the choice of fitting functions.
While the exponential nature of disk profiles has clearly been established
\cite[]{Freeman70}, the true shape of bulge profiles in late-type spirals
appears to have eluded the majority of workers in the field.  Data now suggest
that late-type bulges would be more accurately represented by an exponential
profile instead of the long-assumed de~Vaucouleurs law.  Historical
developments are discussed in \cite{CouJon+96} and \cite{Courteau96a}.

To test for the shape of bulge luminosity profiles, we adopt the formulation
of \cite{Sersic68} who showed that exponential and de Vaucouleurs profiles are
special cases of the general power law:
\begin{equation}
\Sigma(r) = \Sigma_{\rm 0} \exp\left[-\left(r/r_{\rm0}\right)^{1/n}\right] \;,
\label{eq_sersic1}
\end{equation}
where $\Sigma_{\rm 0}$ is the central brightness, $r_{\rm 0}$ is a scaling
radius, and the exponent $1/n$ is a free variable.  $n=1$ corresponds to an
exponential profile and the $n=4$ case is equivalent to a de~Vaucouleurs law.

\begin{figure}
\plottwo{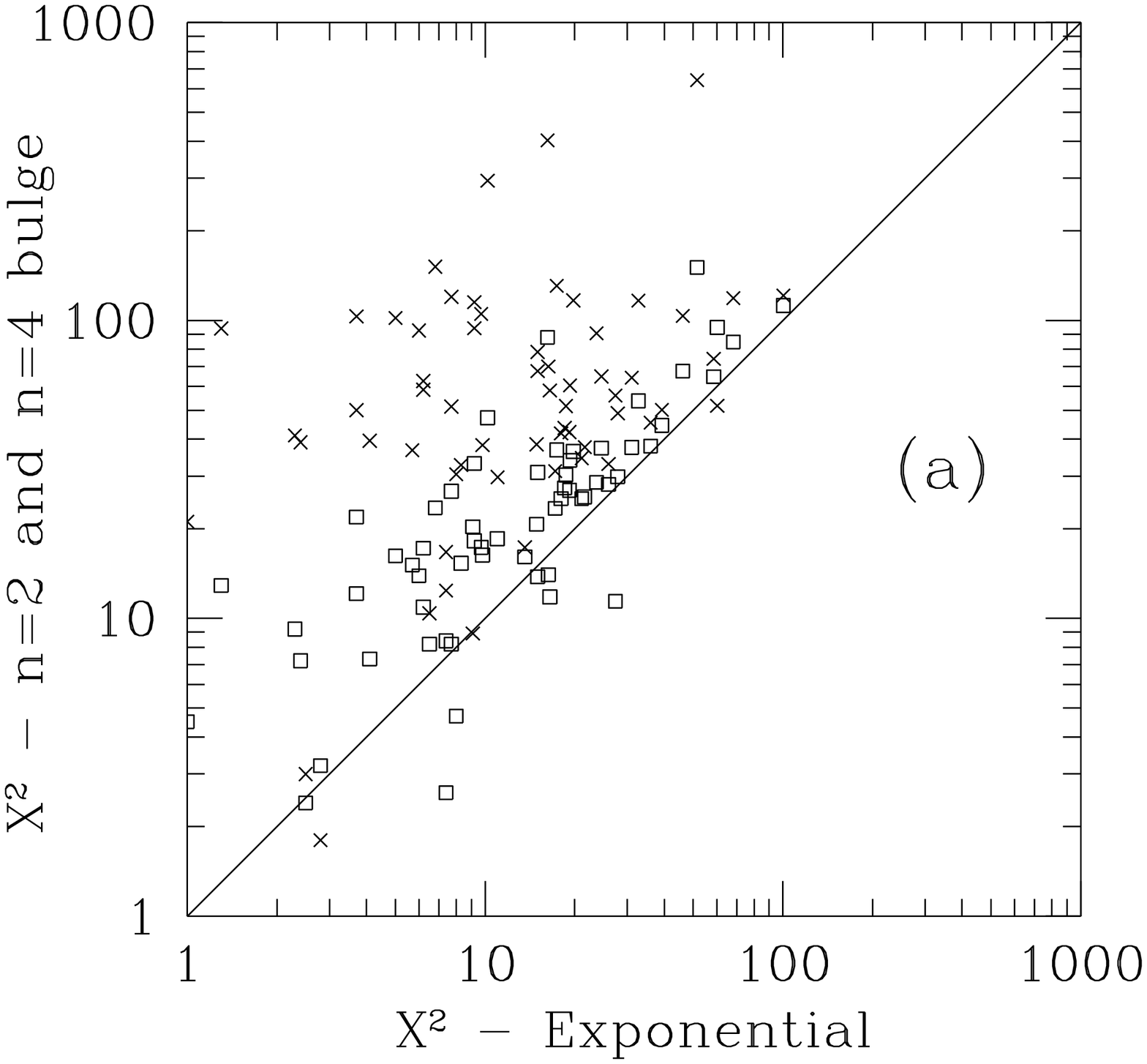}{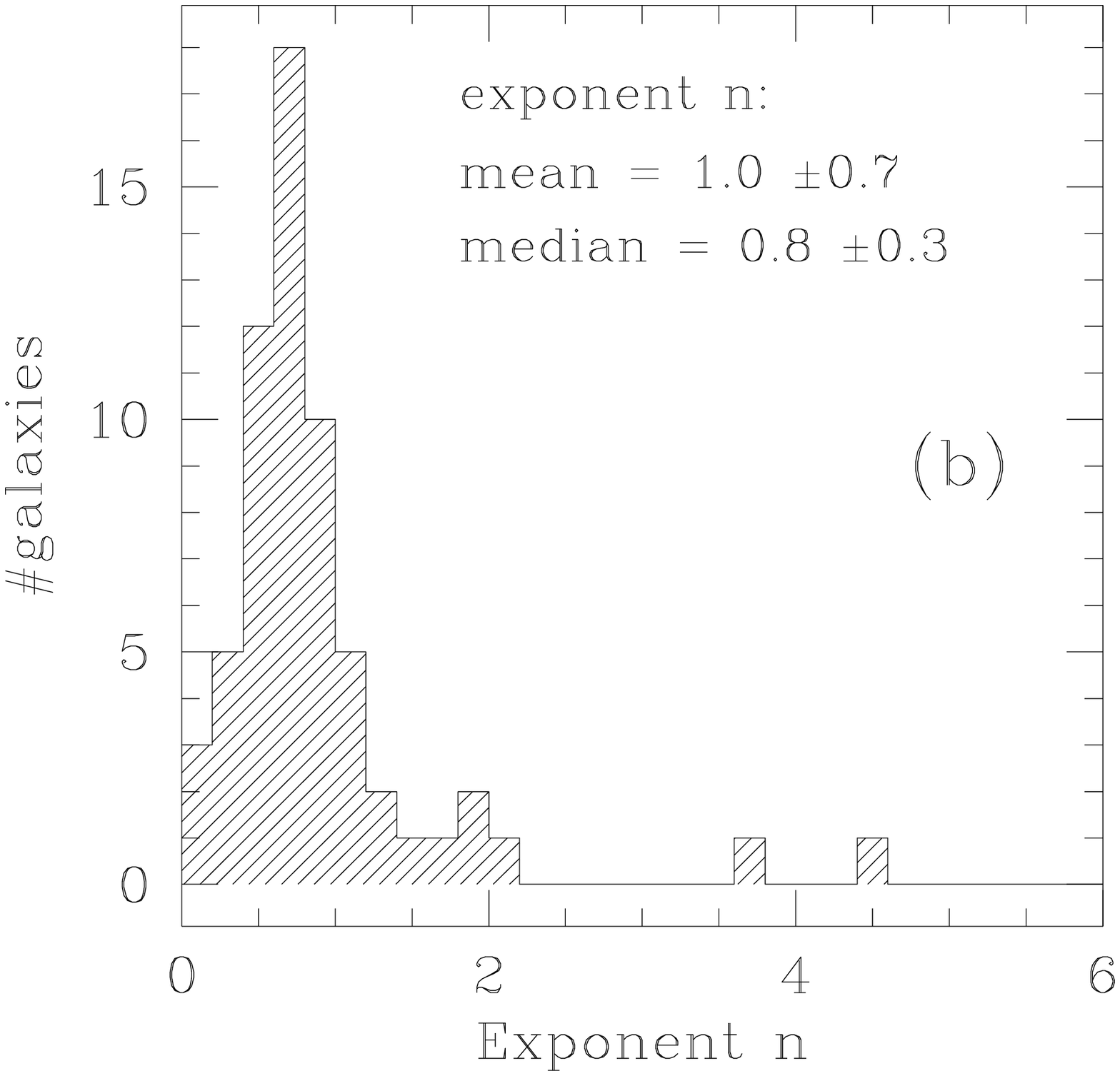}
\caption{($a$) Comparison of $\chi^2$ values for models with $n = 2$ (squares) and 
4 (crosses) with exponential bulge models for the SB profiles of the
``clean'' sub-sample. ($b$) Histogram of the exponent $n$ with
the lowest $\chi^2$ in the fit.
\label{fig_sub_n_chi} }
\end{figure}

\subsection{Double exponential decompositions}
Because many SB profiles often show unevenness (bumps and wiggles), it is not
always possible to determine the best value of $n$ reliably for most galaxies
in our sample. These irregularities are often explained by the presence of
regions of enhanced star formation, spiral arms, bars or other asymmetries in
the 2-D light distribution.  A significant fraction of SB profiles are of
Freeman type II, defined by \cite{Freeman70} as a profile with a dip (near the
bulge/disk transition) with respect to the exponential fit from the outer
disk.  Indeed, our sample is about equally divided between Freeman types I and
II.  In order to establish the best fitting bulge function for the late-type
systems, we first selected galaxies with smooth Freeman type I SB profiles and
with a noticeable central (bulge-like) component.  A total of 63 galaxies were
selected for a combination of 90 independently-determined SB profiles.

Even with this ``cleaner'' sub-sample, solving for all five parameters
($\mud$, $\rd$, $\mub$, $\rb$, and $n$) simultaneously would be fraught with
danger.  Due to the paucity of data points at the center, the simultaneous
solution of three bulge parameters is poorly constrained at best.  Instead, we
solve for a range of {\it fixed} values of $n$ with the other four parameters
unconstrained.  An acceptable solution was usually reached within 10 or fewer
iterations.

The upper panels of Fig.~\ref{fig_sub_sb_rc} show examples of bulge + disk
decompositions with values of $n = 1,\ 2,\ 4$ for UGC~9097, a strong-bulged
spiral by late-type standards.  The quality of the match is usually best
judged in the transition region between the bulge and inner disk.  The failure 
of the $r^{1/4}$ profile is quite noticeable.  
Indeed, for nearly all the SB profiles, exponential
bulges yield the lowest $\chi^2$ values.  This is demonstrated in
Fig.~\ref{fig_sub_n_chi}a which shows the distribution of $\chi^2$'s for $n =
2$ (squares) and 4 (crosses) against the one for exponential bulges.

We sampled values of $n$ between 0.2 and 6 in steps of 0.2, and determined for
each profile the ``best'' value of $n$ corresponding to the lowest $\chi^2$.
The distribution of ``best'' $n$ values, shown in Fig.~\ref{fig_sub_n_chi}b,
peaks at a value of $n = 0.8$, i.e. nearly exponential profiles.  The ratio of
bulge-to-disk exponential scale lengths (not shown here) is close to 0.1, as
predicted by N-body simulations of secular evolution models (Courteau \etal
1996, Courteau 1996b).  In agreement with other recent studies (see
e.g. Courteau \etal 1996, and references therein), we conclude from this that
the central regions of late-type spirals are best fit by exponential profiles,
rather than the de Vaucouleurs law or an $n = 2$ bulge.  For the remainder of
this paper, SB profiles will only be modeled as a combination of two
exponentials for the bulge and disk respectively.

\section{Mass Decompositions}

Modeling of the observed rotation curves follows the non-linear least-squares
procedure outlined by \cite{Kent86} and \cite{Broeils92b} (see also review by
van Albada in this volume). The basic assumption is that the gravitational
potential of a galaxy can be described as the sum of potentials of individual
mass components, like stellar bulge and disk, gas disk, dark halo, etc. In all
cases, the radial distribution of visible mass is assumed to be given by the
mean radial distribution of light obtained from the B/D decompositions of the
SB profiles.  The bulge and disk $M/L$'s are allowed to vary independently but
remain constant (as a function of radius) within each component.  The observed
RCs are well fitted by the luminous components alone. These data provide
little support for a dark component, and therefore its contribution has not
been included in the mass models. This also means that the $M/L$'s derived in
this way are true ``maximum'' disk (and bulge) values, since inclusion of gas
and dark matter will only lower these values.

\begin{figure}
\plotahb{5.25in}{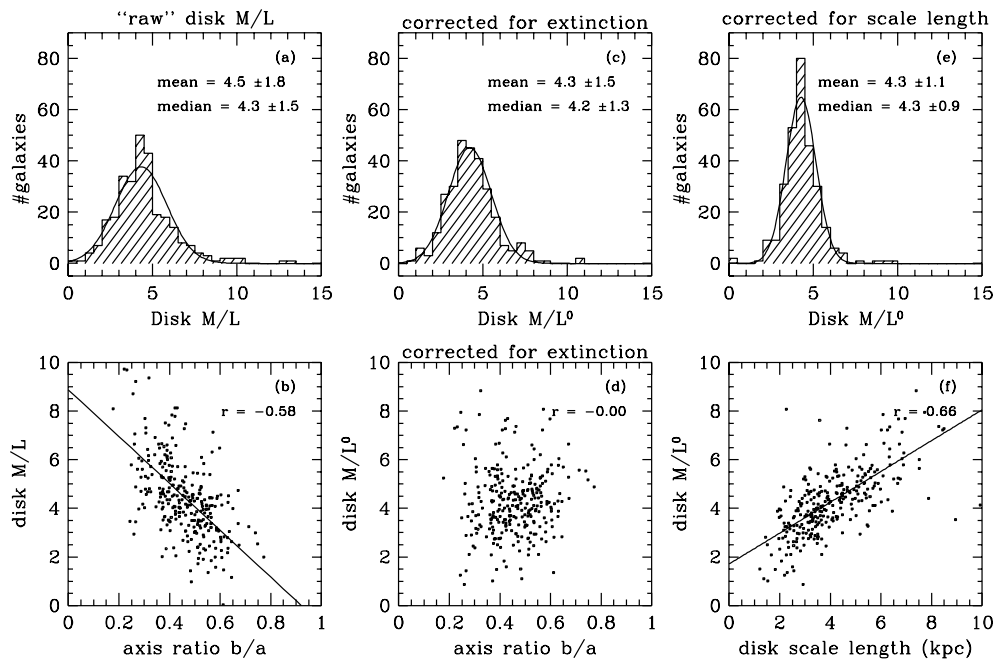}
\caption{($a$) Histogram of ``raw'' disk $M/L$, ($b$) dependence of $M/L$ on
axis ratio $b/a$, ($c$) histogram of ``face-on'' $M/L$, ($d$) $M/L$ after
$b/a$ correction, ($e$) histogram of $M/L^0$ corrected for ``size'' using $h$,
($f$) dependence of $M/L^0$ on $h$.
\label{fig_mlhist} }
\end{figure}

The model disk and bulge rotation velocities are sampled at the same radii as
the observed rotation curve, added in quadrature using the $M/L$'s as free
scaling parameters, and fitted to the observed rotation curve.  All fits were
visually inspected and occasionally the bulge contribution was adjusted
manually ($M/L_{\rm bulge}$ set to maximum allowed value), especially in cases
where the number of data points in the central region was insufficient to
constrain the fits.

\begin{figure}
\plotahb{4.9in}{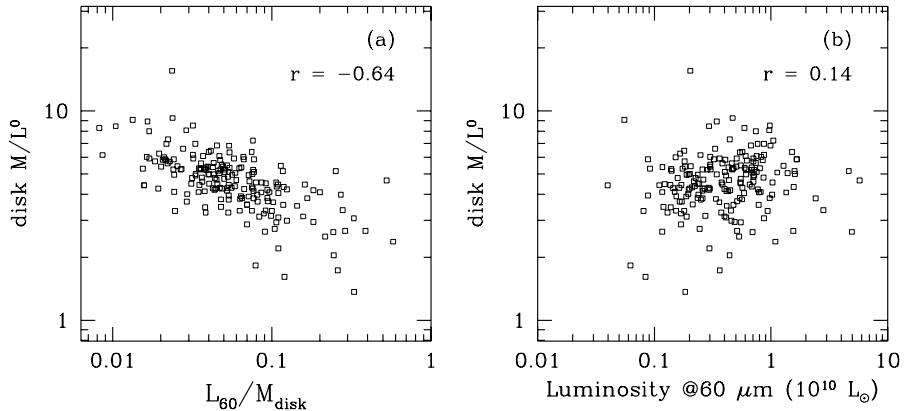}
\caption{($a$) Disk $M/L$ versus $L_{60}/M$. $L_{60}/M$ is proportional to the
SFR per unit mass. ($b$) Disk $M/L$ versus $L_{60}$. No apparent trend is
visible.  The correlation coefficients are shown in the  upper right-hand
corners. 
\label{fig_ml_l60} }
\end{figure}

\subsection{Results}\label{sec_ml}

We shall now examine the variation of $M/L$ (from here onwards, $M/L$ will
refer to {\it disk} $M/L$) among galaxies and try to understand the processes
causing this variation.  Fig.~\ref{fig_mlhist}a shows the distribution of the
``raw'' $M/L$ from the mass decomposition of the rotation curves.  The
$r$-band luminosities used to estimate $M/L$ were corrected for Galactic
extinction only, but these (and consequently the $M/L$'s) need to be corrected
for internal extinction as well.  Unfortunately, there is no trivial and
accepted way to do this due to our very limited understanding of radiation
transfer and sources of opacity in galaxies \cite[e.g. see conf. proc. on this
subject,][]{DavBur95}.  The inclination dependence of $M/L$'s was removed
empirically by fitting the observed trend against ellipticities.
Figure~\ref{fig_mlhist}b shows the dependence of the ``raw'' $M/L$'s on
ellipticity; the line indicates the fit used to correct to ``face-on'' ratios.

Application of this correction reduces the width of the $M/L$ distribution by
17\% (see Fig.~\ref{fig_mlhist}c,d).  The corrected values still show a fairly
broad distribution given the narrow range of galaxy types (Sb--Sc).  One can
appreciate that the mass in these $M/L$'s is dominated by the old disk
population, whereas a larger population of stars contributes to the $r$-band
galaxian flux.  Indeed, \cite{RheAlb96} recently argued that a large part of
the $M/L$ scatter is due to variations in star formation rates (SFRs) among
galaxies. These authors used $L_{60}/M$ as an indicator of the present SFR
per unit mass (their $M$ comes from a maximum-disk fit
analogous to ours), to expose a clear correlation between $\log(M/L)$ and
$\log(L_{60}/M)$ using the Rubin/Kent sample \cite*[]{RubBur+85,Kent86}.
Galaxies with a prominent young population have low $M/L$'s which, they infer,
would be due to enhanced star formation.  This trend is also visible in our
sample as shown in Fig.~\ref{fig_ml_l60}a.  However the lack of correlation
between $M/L$ and $L_{60}$ (a more direct indicator of the SFR) shown in
Fig.~\ref{fig_ml_l60}b suggests that the above effect is most likely due to a
correlation between $M/L$ and the mass of the stellar disk, contrary to the
interpretation of \cite{RheAlb96}.

It is not clear if the mass of the stellar disk itself is the fundamental
physical parameter, but one could argue that the {\it ``size''} of the galaxy
is an important scaling parameter. The ``size'' of a galaxy could be measured
by its (disk) mass, luminosity, linear size or disk scale length ($h$) since
these parameters are all tightly coupled via the Tully-Fisher relation.  The
dependence of $M/L$'s on the scaling parameter $h$ is shown in
Fig.~\ref{fig_mlhist}f.  Compact galaxies have low $M/L$'s and big galaxies
have large $M/L$'s. If we take out this scaling effect using the fitted line
in Fig.~\ref{fig_mlhist}f, we obtain the histogram plotted in
panel~\ref{fig_mlhist}e.  The
spread in $M/L$ can now be completely accounted for by a random error of about
15\% in the distances to these galaxies.  Similar projected distributions can
be obtained by correcting the $M/L$'s for disk mass and luminosity
\cite[]{BroCou97}.

\begin{figure}
\plotahb{6cm}{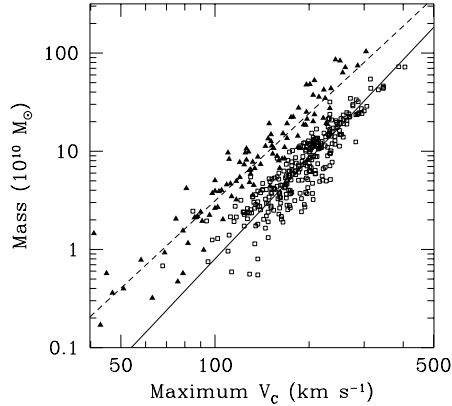}
\caption{Mass versus maximum rotation velocity. The open symbols indicate
luminous mass, filled triangles indicate total mass obtained from
\protect\cite{BroRhe97}.  The figure indicates that luminous mass fraction
(solid line) increases with size.
\label{fig_convergence} }
\end{figure}

Finally, we combine the results from our mass models with those from
\ion{H}{I} data on separate galaxies (Broeils \& Rhee 1997) to show that the
dark matter fraction in galaxies, expressed as the dark-to-luminous mass
fraction $M_{\rm dark}/M_{\rm lum}$, also depends on a structural (``size'')
parameter.  The optical RCs themselves do not extend far enough to allow for
the detection of DM.  However, they provide a good estimate of the total
luminous mass\footnote{So long as the mass model fits indicate that the amount
of dark matter in regions sampled by the optical RCs is small.}  and of the
dependence of this mass on ``size''.  The open squares in
Fig.~\ref{fig_convergence} indicate the {\it luminous} mass of the 290
galaxies used in this study, and the full line is a linear fit to these data
points.  In lieu of scalelengths, the maximum rotation velocity ($V_{\rm max}$) 
acts as the ``size'' parameter.  The filled triangles represent {\it total} masses
obtained from 1-D \ion{H}{I} spectra of 103 galaxies \cite[]{BroRhe97} (fitted
with a dashed line).  It is obvious that the luminous mass fraction (with
respect to the total mass) increases systematically with bigger galaxy size.
Thus, for arbitrary large systems and if the maximum-disk hypothesis holds
true, the amount of dark matter per galaxy becomes negligible.  The increase
of luminous mass fraction with galaxy size has been known for quite some time
based on optical RCs alone \cite*[]{PerSal88,PerSal90c} and \ion{H}{I} RCs
\cite[][and references therein]{Broeils92b}; the presentation here shows the
trend for much larger and completely independent samples.  While the slope
from the \ion{H}{I} samples is fixed, that of our optical decompositions
depends highly on the assumption of maximum-disk (as in the case of detailed
mass models for extended RCs).  A more important contribution of the dark
matter to the inner disk would cause the trend to flatten.

\section{Conclusions}
We find that the luminosity distributions of bulges of Sb--Sc galaxies are
best fitted by exponential profiles. Using such profiles, we find that the
ratio of bulge-to-disk scale lengths is close to 0.1, as predicted by
simulations of secular evolution. Disk $M/L$'s obtained from maximum-disk fits
to optical RCs correlate well with galaxy ``size''.  Correcting for this
``size'' effect yields a narrow distribution of intrinsic $M/L$'s, which can
be fully accounted for by 15\% distance errors.

\end{document}